\documentclass[prl,twocolumn,twoside,preprintnumbers,superscriptaddress,nofootinbib]{revtex4}
\usepackage{amsmath}
\usepackage{amssymb}
\usepackage{hyperref}
\usepackage{xspace}
\usepackage{slashed}
\hypersetup{
    colorlinks=true,       
    linkcolor=blue,          
    citecolor=blue,        
    filecolor=blue,      
    urlcolor=blue           
}
\usepackage{xcolor}
\usepackage{color}
\usepackage{graphicx}  %
\usepackage{bm}  %
\usepackage{amsmath}   %
\usepackage{graphics}

\newcommand{\al}{\alpha}

\newcommand{\g}{\gamma}

\newcommand{\MQCD}{\Lambda_\chi}

\newcommand{\Or}{\mathcal O}

\newcommand{\vL}{\ensuremath{\mathcal{L}}}
\newcommand{\vp}{\varphi}
\newcommand{\sq}{^{2}}

\newcommand{\dslash}[1]{#1 \llap{/\kern-0.5pt}}
\newcommand{\Dslash}[1]{#1 \llap{/\kern+1.5pt}}
\newcommand{\DDslash}[1]{#1 \llap{/\kern+2.3pt}}
\newcommand{\dslashh}[1]{#1 \llap{/\kern+1pt}}

\newcommand{\boldtau}{\mbox{\boldmath $\tau$}}
\newcommand{\boldpi}{\mbox{\boldmath $\pi$}}

\newcommand{\CP}{$C\!P$}
\newcommand{\CPspace}{$C\!P\!$ }
\newcommand{\CPbf}{$\boldsymbol{C\!P}\!$ }
\newcommand{\Ex}[1]{\cdot 10^{#1}}
\newcommand{\bea}{\begin{eqnarray}}
\newcommand{\eea}{\end{eqnarray}}
\newcommand{\be}{\begin{equation}}
\newcommand{\ee}{\end{equation}}
\newcommand{\bma}{\begin{pmatrix}}
\newcommand{\ema}{\end{pmatrix}}
\newcommand{\nn}{\nonumber}

\newcommand{\eps}{\epsilon}
\newcommand{\GeV}{\,\text{GeV}}
\newcommand{\Order}{\mathcal O}

\newcommand{\dHg}{d_{\rm Hg}}
\newcommand{\dXe}{d_{\rm Xe}}
\newcommand{\dRa}{d_{\rm Ra}}

\allowdisplaybreaks[1]

\begin{document}
\preprint{INT-PUB-19-008, LA-UR-19-22027, PSI-PR-19-01, ZU-TH 08/19,  RBRC-1316}

\title{
\CPbf violation in Higgs--gauge interactions: from tabletop experiments to the LHC
}

\author{Vincenzo Cirigliano}
\affiliation{Theoretical Division, Los Alamos National Laboratory, Los Alamos, NM 87545, USA}

\author{Andreas Crivellin}
\affiliation{Paul Scherrer Institut, CH--5232 Villigen PSI, Switzerland}
\affiliation{Physik-Institut, Universit\"at Z\"urich, Winterthurerstrasse 190, CH--8057 Z\"urich, Switzerland}

\author{Wouter  Dekens}
\affiliation{Department of Physics, University of California at San Diego, La Jolla, CA 92093, USA} 

\author{Jordy de Vries}
\affiliation{Amherst Center for Fundamental Interactions, Department of Physics, University of Massachusetts, Amherst, MA 01003}
\affiliation{RIKEN BNL Research Center, Brookhaven National Laboratory,
Upton, New York 11973-5000, USA}

\author{Martin Hoferichter}
\affiliation{Institute for Nuclear Theory, University of Washington, Seattle, WA 98195-1550, USA}

\author{Emanuele Mereghetti}
\affiliation{Theoretical Division, Los Alamos National Laboratory, Los Alamos, NM 87545, USA}
 
\begin{abstract}
We investigate the interplay between the high- and low-energy phenomenology of \CP-violating interactions of the Higgs boson with gauge bosons. For this purpose we use
an effective field theory approach and consider all dimension-6 operators arising in so-called universal theories.  
We compute their loop-induced contributions to electric dipole moments and the \CPspace asymmetry in $B\to X_s\gamma$,
and compare the resulting current and prospective constraints to the projected sensitivity of the LHC.
Low-energy measurements are
shown to generally have a far stronger constraining power, which results in highly correlated allowed
regions in coupling space---a distinctive pattern that could be probed at the high-luminosity LHC.
\end{abstract}

\maketitle

{\it Introduction.}---To generate the observed matter--antimatter asymmetry in the Universe the Sakharov conditions~\cite{Sakharov:1967dj} have to be satisfied. One of them  requires that charge-parity (\CP) symmetry be violated. \CPspace symmetry is broken in the Standard Model (SM) of particle physics with three generations of quarks, but only by the phase of the Cabibbo--Kobayashi--Maskawa (CKM) matrix and, potentially, the QCD $\theta$ term. The resulting amount of \CPspace violation is, however, far too small to explain the observed matter--antimatter asymmetry~\cite{Cohen:1993nk,Gavela:1993ts,Huet:1994jb,Gavela:1994ds,Gavela:1994dt,Riotto:1999yt}. Scenarios of electroweak (EW) baryogenesis~\cite{Kuzmin:1985mm,Shaposhnikov:1987tw,Nelson:1991ab,Morrissey:2012db} demand new sources of \CPspace violation not too far above the EW scale.

It has long been recognized that the required new \CP-violating couplings 
can generate observable effects in  both   Higgs production and decay rates, i.e.\ \CP-even observables
\cite{Pospelov:2005pr,Li:2010ax,McKeen:2012av,Harnik:2012pb,Engel:2013lsa,Shu:2013uua,Chen:2014gka,Inoue:2014nva,Dwivedi:2015nta,Chien:2015xha,Cirigliano:2016njn,Cirigliano:2016nyn,Dekens:2018bci},   
as well as 
genuinely \CP-odd signatures   at the Large Hadron Collider
 (LHC)~\cite{Soni:1993jc,Plehn:2001nj,Ferreira:2016jea,Dawson:2013bba,Anderson:2013afp,Bernlochner:2018opw,Alioli:2017ces,Alioli:2018ljm,Sirunyan:2019twz,Sirunyan:2019nbs,Aaboud:2017fye,Hirschi:2018etq,Englert:2019xhk}.     Moreover,  the interplay with low-energy \CP-violating observables such as electric dipole moments (EDMs) has been  explored, 
in either specific models~\cite{Bian:2014zka,Inoue:2014nva} or SM effective field theory (SMEFT)~\cite{Pospelov:2005pr,Li:2010ax,McKeen:2012av,Engel:2013lsa,Inoue:2014nva,Dwivedi:2015nta,Chien:2015xha,Cirigliano:2016njn,Cirigliano:2016nyn,Dekens:2018bci},   
taking into account only subsets of  the dimension-6 \CP-odd operators.  

Here we take a novel point of view and focus on  the \CP-violating sector of  so-called universal theories,   
originally introduced as the broad class of SM extensions in which beyond-the-SM
(BSM) particles   couple  to  SM  bosons and/or  to SM fermions only through the  gauge and Yukawa currents~\cite{Barbieri:2004qk},   
placing the analysis of  the oblique EW corrections~\cite{Kennedy:1988sn,Peskin:1990zt}  
and EW precision tests  (EWPTs) in a more general  and consistent framework. 

With the forthcoming   high-luminosity (HL) LHC  upgrade, EWPTs  
involving triple-gauge-boson and gauge-boson--Higgs couplings 
will be an important thrust, 
and will  probe universal theories beyond the level reached with LEP/SLC data~\cite{Cepeda:2019klc,Azzi:2019yne,deBlas:2019rxi}.  
In this context,  both for completeness and in connection with baryogenesis, 
it is timely to study the \CP-violating sector of such theories and to investigate {\it  quantitatively} the complementarity 
of collider and low-energy measurements.

To address this problem we work within SMEFT, which relies on assuming a gap 
between the scale $\Lambda$ of BSM physics and the EW scale. 
Universal theories induce, modulo field redefinitions,  only bosonic operators  at the scale $\Lambda$~\cite{Wells:2015uba}. 
The SMEFT setup for the \CP-conserving sector of universal theories and the effect of  non-universal operators 
 generated by the renormalization group (RG)  flow have been studied in Refs.~\cite{Wells:2015uba,Wells:2015cre}.    
We find that the \CP-violating sector of universal theories is characterized by  six  dimension-6  
operators,  which in the Warsaw basis~\cite{Buchmuller:1985jz,Grzadkowski:2010es}  read 
\begin{align}
\label{HiggsGauge}
\mathcal L &=
 -g^2 C_{\varphi  \tilde W} \, \varphi^\dagger \varphi  \, \tilde W^i_{\mu \nu} W^{\mu \nu}_i 
 -g'^2 C_{\varphi \tilde B} \, \varphi^\dagger \varphi     \,  \tilde B_{\mu \nu} B^{\mu \nu}
 \\
&- g g' C_{\varphi \tilde{W} B} \, \varphi^\dagger \tau^i  \varphi     \,  \tilde W^i_{\mu \nu}  B^ {\mu \nu}
 - g_s^2 C_{\varphi  \tilde G}\, \varphi^\dagger \varphi    G^a_{\mu \nu} \tilde G^{\mu \nu}_a
\notag
\\
& +  \frac{C_{\tilde G}}{3} \, g_sf_{abc} \tilde G_{\mu\nu}^a G^{\nu\rho}_b G^{c\,\mu}_\rho  + 
\frac{C_{\tilde W}}{3}  \   g\eps_{ijk} \tilde W_{\mu\nu}^i W^{\nu\rho}_j W^{k\,\mu}_\rho \,, 
\notag
\end{align}
where $\varphi$ is the Higgs doublet with $\left\langle \varphi \right\rangle  = v/\sqrt 2$, $v\simeq 246\GeV$, 
$g_s$, $g$, and $g'$ are the $SU(3)_c$, $SU(2)_L$, and $U(1)_Y$ couplings, respectively, and $G_{\mu\nu}$, $W_{\mu\nu}$, and $B_{\mu\nu}$ the corresponding field strength tensors. 
We define  $\tilde X^{\mu\nu} = \eps^{\mu\nu\alpha\beta} X_{\alpha\beta}/2$, $\eps^{0123}=+1$. 
The Wilson coefficients $C_{\varphi \, \tilde X, \tilde{X}}$  encode contributions from BSM physics scaling as $1/\Lambda^2$.

This scenario has additional desirable features:
it provides   a natural arena to study   \CP-violating Higgs--gauge interactions in the SMEFT context, as those 
 arise, together with the triple-gauge-boson, as the dominant \CP-violating couplings. Furthermore,
the BSM scale  $\Lambda$ can be relatively low (as minimal flavor violation~\cite{DAmbrosio:2002vsn,Cirigliano:2005ck} 
is satisfied and  \CP-violating fermionic  dipoles are generated   only through RG flow),  
a welcome feature for  the viability of  weak-scale baryogenesis.

The  operators in Eq.~\eqref{HiggsGauge} affect the cross sections of processes such as Higgs production via gluon or vector-boson fusion, Higgs production in association with EW gauge bosons, and Higgs decays, through non-interfering contributions quadratic in $C_{\varphi \tilde X}$ and are thus suppressed by $(v/\Lambda)^4$. Such dimension-8 contributions however still lead to significant constraints~\cite{Ferreira:2016jea,Alioli:2018ljm}. The Higgs--gauge operators contribute at $\Order(v^2/\Lambda^2)$ to \CP-odd observables, such as the \CPspace asymmetry in $p p \rightarrow h + 2 j$~\cite{Dawson:2013bba,Anderson:2013afp,Bernlochner:2018opw,Sirunyan:2019nbs}, angular distributions in associated $HW$ and $HZ$ production~\cite{Ferreira:2016jea,Alioli:2017ces,Alioli:2018ljm}, or in $h \rightarrow 4 l$~\cite{Soni:1993jc,Sirunyan:2019twz,Sirunyan:2019nbs}, while $C_{\tilde W}$ and $C_{\varphi \tilde W B}$
contribute to \CP-odd observables in diboson production~\cite{Ferreira:2016jea,Aaboud:2017fye}. 
$C_{\tilde G}$ gives tree-level corrections to $p p \rightarrow h + 2 j$ and to multijet production~\cite{Hirschi:2018etq}. In addition to these tree-level effects in collider observables, 
all coefficients contribute to 
low-energy \CP-violating observables,  such as  EDMs and the  \CPspace asymmetry in $B\to X_s\gamma$,   at the loop level. In this Letter   we set up the framework to include  low-energy  \CP-violating  probes  and demonstrate that they put severe constraints on 
the  \CP-violating sector of  universal theories. To establish the connection to existing collider bounds~\cite{Bernlochner:2018opw,Englert:2019xhk}, we first concentrate the phenomenological analysis on the operators that  involve the Higgs coupling, 
and later discuss the low- and high-energy  input necessary 
for an  analysis of all six parameters simultaneously.

\begin{figure}
\includegraphics[width=\linewidth]{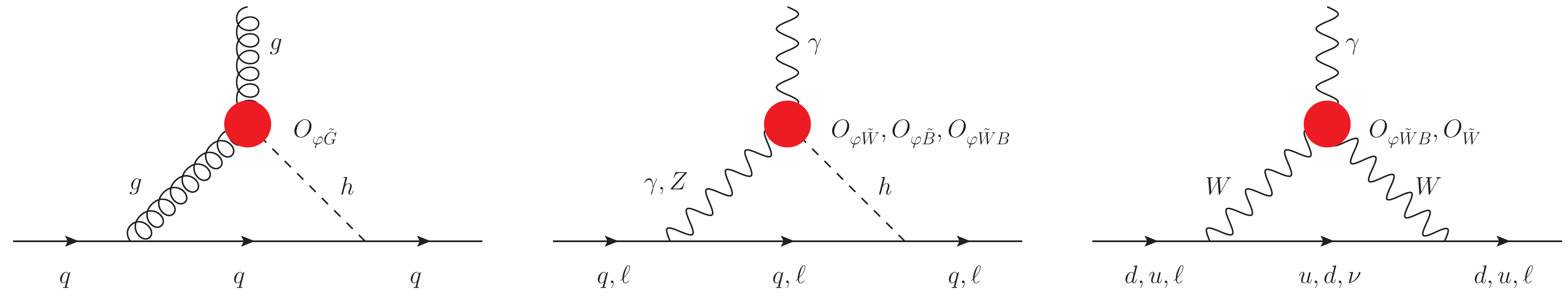}
\caption{One-loop diagrams involving Higgs--gauge operators that contribute to (gluonic) dipole operators. The red circles denote insertions of the SMEFT operators. 
The diagram on the right side also generates threshold corrections to flavor-violating dipole operators.
}
\label{diagrams}
\end{figure}

{\it Renormalization group evolution.}---When the Higgs field acquires its vacuum expectation value, the operators in Eq.~\eqref{HiggsGauge} generate $\theta$-like terms
by means of $\varphi^\dagger \varphi\to v^2/2 + \dots$, $\varphi^\dagger\tau^i \varphi\to -\delta^{i3} v^2/2 + \dots$, where the dots denote terms that contain the Higgs scalar boson $h$.  
The parts of the operators in Eq.~\eqref{HiggsGauge} that do not involve $h$ can be absorbed in the SM $\theta$ terms. The $U(1)_Y$ and $SU(2)_L$ $\theta$ terms are unphysical because they can be removed by field rotations~\cite{Anselm:1992yz,Anselm:1993uj,Perez:2014fja}. 
The gluonic operator effectively shifts the QCD $\theta$ term
$\theta \rightarrow \theta - 16\pi^2 v^2 C_{\varphi \, \tilde  G}$, which is strongly constrained by the neutron EDM~\cite{Baker:2006ts,Afach:2015sja}. However, we will assume the presence of a Peccei--Quinn (PQ) mechanism~\cite{Peccei:1977hh} under which the total $\theta$ term vanishes dynamically.

Below the EW scale, the Lagrangian contains flavor-conserving operators that induce leptonic and hadronic EDMs (fermion EDMs, quark chromo EDMs (CEDMs), and the Weinberg operator) as well as $\Delta B =\Delta S= 1$ operators that contribute to $B \rightarrow X_s \gamma$,
through the diagrams shown in Fig.~\ref{diagrams}. These diagrams provide both finite matching contributions at the EW scale, $\mu=\mu_t$, and contributions to the anomalous dimensions that determine the RG evolution between the BSM scale, $\mu=\Lambda$, and the EW scale. We then evolve the low-energy operators to the scale where QCD becomes non-perturbative, $\mu=\MQCD= 2$ GeV, and take into account the bottom, charm, and strange thresholds where additional matching contributions are generated. More details about the evolution from the high- to low-energy scale are given in Ref.~\cite{SuppRG} 
(including Refs.~\cite{Bobeth:2015zqa,He:1993hx,Dekens:2013zca,Alonso:2013hga,Chanowitz:1979zu,tHooft:1972tcz,Weinberg:1989dx,Wilczek:1976ry,Braaten:1990gq,Degrassi:2005zd}).

A key outcome of the RG analysis is that the weak operators $C_{\varphi \tilde B}$, $C_{\varphi \tilde W}$, $C_{\varphi \tilde W B}$, and $C_{\tilde W}$ 
contribute to the fermion EDMs almost exclusively via two combinations,
proportional to the third component of the weak isospin, $T^3_f$, and the electric charge, $Q_f$. For this reason, present and future EDM experiments constrain at most four directions in the parameter 
space of Eq.~\eqref{HiggsGauge}, up to small subleading effects.

{\it Low-energy observables.}---Next, we discuss the  connection to the most sensitive low-energy observables, starting with EDMs. The most stringent limits are set by the neutron and ${}^{199}$Hg atom, and by measurements on the polar molecule ThO. For the operators in Eq.~\eqref{HiggsGauge}, the ThO measurement~\cite{Andreev:2018ayy,Baron:2013eja} can be interpreted as a probe of the electron EDM, with a small theoretical uncertainty~\cite{Skripnikov,Fleig:2014uaa}. In contrast, nucleon, nuclear, and diamagnetic EDMs receive contributions from several operators, with varying levels of theoretical uncertainties. We provide the full expressions in Ref.~\cite{SuppLE} (including Refs.~\cite{Gupta:2018lvp, Alexandrou:2017qyt,Pospelov:2000bw,Lebedev:2004va,Hisano:2012sc,Demir:2002gg,deVries:2012ab,Pospelov:2001ys,Bsaisou:2014zwa,Dmitriev:2003sc,deJesus:2005nb,Ban:2010ea,Dzuba:2009kn,Sahoo:2016zvr,Fleig:2018bsf,Yamanaka:2017mef,Yanase:2018qqq,Dobaczewski:2018nim,Benzke:2010tq}).

Matrix elements connecting quark EDMs to nucleon EDMs are relatively well known~\cite{Gupta:2018lvp}, but contributions from quark CEDMs and the Weinberg operator suffer from larger uncertainties. In addition to nucleon EDMs, nuclear and diamagnetic EDMs are generated by \CP-odd nuclear forces that, for the operators under consideration, are dominated by \CP-odd one-pion exchange between nucleons. The sizes of the associated low-energy constants have been calculated with QCD sum rules~\cite{Pospelov:2001ys}, with $\Or(100\%)$ hadronic uncertainty. In addition, the nuclear many-body matrix elements that determine diamagnetic EDMs involve sizable nuclear uncertainties. 

\begin{table}[t]
\small
\centering
\begin{tabular}{ccccc}
\toprule
$d_e$ & $d_n$& $\dHg$  & $\dXe$ & $\dRa$\\
$1.1 \cdot 10^{-29} $ &$ 3.0 \cdot 10^{-26}$  & $6.2 \cdot 10^{-30}$  & $3.9 \cdot 10^{-27}$ &$1.2\cdot 10^{-23}$\\
\botrule
\end{tabular}
\caption{Current limits on the electron~\cite{Andreev:2018ayy}, neutron~\cite{Baker:2006ts,Afach:2015sja}, mercury~\cite{Griffith:2009zz,Graner:2016ses},
xenon~\cite{PhysRevLett.86.22,Sachdeva:2019rkt}, and radium~\cite{Parker:2015yka,Bishof:2016uqx} EDMs in units of $e$ cm ($90\%$ C.L.).
The result for the \CPspace asymmetry, $A_{B\to X_s\gamma}=0.015(20)$, is taken from Refs.~\cite{Nishida:2003paa,Lees:2014uoa,Amhis:2016xyh}.}
\label{tab:EDMexps}  
\end{table}

Current experimental limits are summarized in Table~\ref{tab:EDMexps}, which also shows the limits on systems that are not yet competitive, but could provide interesting constraints in the future. EDM experiments on ${}^{225}$Ra and ${}^{129}$Xe atoms have already provided limits~\cite{PhysRevLett.86.22, Parker:2015yka,Sachdeva:2019rkt} and are quickly improving. 
Plans exist to measure the EDMs of charged nuclei such as the proton and deuteron in electromagnetic storage rings~\cite{Eversmann:2015jnk}. The EDM measurements of light nuclei can be more reliably interpreted in terms of BSM operators than is the case for $d_{\rm Hg}$ as the nuclear theory is under solid theoretical control~\cite{deVries:2011an,Bsaisou:2014zwa}.

The operators $O_{\tilde W}$ and $O_{\vp \tilde WB}$  contribute to the \CPspace asymmetry in $B\rightarrow X_s \gamma$ and to \CP-odd triple-gauge couplings that were probed at LEP. Concerning the $B\to X_s\gamma$ asymmetry, we employ the expressions derived in Ref.~\cite{Benzke:2010tq} and take the required SM Wilson coefficients, as well as the hadronic parameters, from the same work. The triple-gauge vertices induced by $O_{\tilde W}$ and $O_{\vp \tilde WB}$ are of the form $W^+W^-\g$ and $W^+W^-Z$, which were constrained using angular distributions in $e^+ e^-\rightarrow W^+ W^-$~\cite{Abbiendi:2000ei,Abdallah:2008sf}. In the notation of Ref.~\cite{Gounaris:1996uw} we have, $\tilde \lambda_Z = \tilde \lambda_\g = -2m_W^2 C_{\tilde W}$ and $\tilde \kappa_Z=-t_w^2 \tilde\kappa_\g = 4t_w^2m_W^2 C_{\vp \tilde WB}$, $t_w=\tan\theta_w$, which leads to~\cite{Tanabashi:2018oca}
\bea\label{LEP}
v^2 C_{\varphi \tilde W B}= -0.93^{+0.47}_{-0.31}\,,\qquad v^2C_{\tilde W} = 0.42(33)\,.
\eea
As shown in Table~\ref{Tab:bounds}, these constraints have already been improved by the study of the $W^+ W^-$ cross section at the LHC~\cite{Aaboud:2019nkz},
and are likely to improve further in the context of EWPTs anticipated at the HL-LHC~\cite{Cepeda:2019klc,Azzi:2019yne,deBlas:2019rxi}.

{
\begin{table}
	\centering
	\scalebox{0.95}{
	\begin{tabular}{c c |c || c}
		\toprule
		 & Central     & Rfit     & LHC  \\
		\colrule
		$v^2\, C_{\varphi \tilde B}$     & $[-5.1,5.1] \cdot 10^{-6}$   & $[-5.1,5.1]\cdot 10^{-6}$   & $[-28,10]$     \\
		$v^2\, C_{\varphi \tilde W}$     & $[-4.7,4.7] \cdot 10^{-6}$ & $[-4.7,4.7]\cdot 10^{-6}$     & $[-2.3,0.43] $   \\
		$v^2\, C_{\varphi \tilde W B}$   & $[-2.2,2.2] \cdot 10^{-6}$& $[-2.2,2.2] \cdot 10^{-6}$    &      $[-0.57,0.57]$   \\ 
		$v^2\, C_{\varphi \tilde G}$ 	 & $[-5.3,5.3] \cdot 10^{-5} $ & $[-1.2,1.2] \cdot 10^{-3} $   &  $[-1.3,8.1]\cdot 10^{-3} $ \\\hline
		$v^2\, C_{\tilde G}$ 	 & $[-2.4,2.4] \cdot 10^{-6}$  & $[-3.4,3.4] \cdot 10^{-5}$  & --  \\		
		$v^2\, C_{\tilde W}$ 	 & $[-4.8,4.8] \cdot 10^{-5}$  &  $[-4.8,4.8]  \cdot 10^{-5}$  & $[-3.1,3.1] \cdot 10^{-2}$\\				
		\botrule
	\end{tabular}
	}
	\caption{
	Central and Rfit low-energy constraints (at $95\%$ C.L.), assuming one of the couplings, $C_\alpha$, is present at the scale $\Lambda = 1$ TeV. 
	For comparison, we show 
	current collider limits from Refs.~\cite{Sirunyan:2019twz,Sirunyan:2019nbs} (for $C_{\varphi \tilde B}$), Refs.~\cite{Aaboud:2017fye,Aaboud:2019nkz} (for  $C_{\varphi \tilde W B}$   and $ C_{\tilde W}$),
	and Ref.~\cite{Bernlochner:2018opw} (for all other couplings). 	
	 }
	\label{Tab:bounds}
\end{table}
}

{\it Analysis.}---To constrain the Higgs--gauge operators, we use EDM limits and the \CPspace asymmetry in $B \rightarrow X_s \gamma$ as listed in Table~\ref{tab:EDMexps}, as well as the LEP constraints on triple-gauge couplings given in Eq.~\eqref{LEP}. Nuclear and hadronic EDMs as well as the \CPspace asymmetry are affected by significant theoretical uncertainties. We follow Ref.~\cite{Cirigliano:2016nyn} and present limits in a variety of cases: (i) the ``central'' scenario, in which we neglect all hadronic and nuclear uncertainties, (ii) the ``Rfit'' strategy, in which all hadronic and nuclear matrix elements are varied within their allowed ranges to minimize the $\chi^2$ value, and (iii) the ``Gaussian'' strategy, in which the theoretical errors are treated in the same way as statistical errors are. 
This last strategy provides a realistic estimate of the impact of the theoretical errors when these are under control. 
We start by discussing the limits derived in the central case, which reflects the maximal constraining power of the low-energy measurements, assuming a single operator is present at the scale $\mu=\Lambda$. We subsequently consider the impact of the theoretical uncertainties in the Rfit scenario, as well as a scenario in which multiple Higgs--gauge operators appear at the scale $\Lambda$.

Turning on a single operator at the scale $\Lambda$, we see from  Table~\ref{Tab:bounds} that the 
low-energy limits are very stringent. The bounds on the operators with EW gauge bosons are dominated by the electron EDM, which constrains  $v^2 C_{\varphi \tilde W,\varphi \tilde B,\varphi \tilde W B,\tilde W}$ to be $ \Or( 10^{-6})$, corresponding to a BSM scale of $\sim 100$ TeV, assuming $C_i=1/\Lambda^2$, or $10$ TeV,  including a loop factor, $C_i = 1/(4\pi\Lambda)^2$. The constraints from the neutron and $^{199}$Hg EDMs are weaker, at the permille level for $v^2 C_{\varphi \tilde W}$ and $v^2 C_{\varphi \tilde W B}$ and at the percent level for $v^2C_{\varphi \tilde B,\tilde W}$. The bounds on $ C_{\varphi \tilde G}$ and $ C_{ \tilde G}$ are dominated by the mercury EDM in the central case. For both operators, the large uncertainties on the matrix element of the Weinberg operator imply that the constraints weaken by an order of magnitude and become dominated by the neutron EDM when moving from the central to the Rfit strategy. In contrast, the limits on the EW operators are very similar when using the Rfit strategy, as they are dominated by the ThO measurement. 
The fourth column in Table~\ref{Tab:bounds} shows the current collider limits
for comparison.\footnote{Here we considered only limits arising from genuine dimension-6 contributions to \CP-violating observables (more information on the CMS limits~\cite{Sirunyan:2019twz,Sirunyan:2019nbs}
is provided in Ref.~\cite{SuppCMS}). Constraints on $v^2 C_{\tilde G}$ stemming from dimension-8 contributions to jet cross sections were considered in Ref.~\cite{Hirschi:2018etq}, and estimated to be $\Or(10^{-2})$.} These high-energy probes are less sensitive by four to six orders of magnitude for most of the couplings, while they are competitive with the EDM constraints on $v^2 C_{\vp \tilde G}$ in the Rfit approach.

\begin{table}
\small
	\centering
	\begin{tabular}{c c c c}
		\toprule
		 & Low energy   & LHC (3000 fb$^{-1}$) \\
		\colrule
		$v^2\, C_{\varphi \tilde B}$     & $[-0.4,0.00]$   & $[-0.3,0.3]$     \\
		$v^2\, C_{\varphi \tilde W}$     & $[-2.3,0.02]$     & $[-0.17,0.17] $   \\
		$v^2\, C_{\varphi \tilde W B}$   & $[-1.3,0.01]$    &  $[-0.39,0.39]$      \\
		$v^2\, C_{\varphi \tilde G}$ 	 & $[-1.3,1.3]\cdot 10^{-5}$   &  $[-9.0,9.0]\cdot 10^{-4} $ \\\hline				
		\botrule
	\end{tabular}\\
	\caption{Comparison of projected  collider and low-energy limits. The LHC limits were taken from Ref.~\cite{Bernlochner:2018opw}, while the low-energy limits assume improved matrix elements and future EDM measurements as described in the text.
	All four couplings were turned on at the scale $\Lambda = 1$ TeV, and the low-energy limits were obtained using the Gaussian strategy for the theoretical uncertainties.}
	\label{Tab:Futbounds}
\end{table}

To see the effects of turning on multiple operators at the scale $\Lambda$, we investigate a scenario in which all Higgs--gauge couplings are present at $\mu=\Lambda$, while keeping $C_{\tilde G, \tilde W}(\Lambda)=0$. This allows us to directly compare the low-energy limits to those of Ref.~\cite{Bernlochner:2018opw}.
In this case there is one free direction left unconstrained by EDM measurements, even when neglecting theoretical uncertainties. For our choice of $\mu_0 = 1$ TeV, this combination of couplings is given by 
$\sim 0.17\, C_{\vp \tilde B}+0.86\,C_{\vp \tilde W}+0.48\,C_{\vp \tilde WB}$. EDM measurements are not sufficient to constrain all four dimension-6 operators simultaneously and the \CPspace asymmetry in $B \rightarrow X_s \gamma$ and LEP observables are needed to close the free direction. 
When treating the theoretical uncertainties in the Rfit or Gaussian approach, the constraints from $d_{\rm Hg}$ and $d_{n}$ are degenerate, leading to another free direction. 
These free directions can be closed by reducing the errors on the theoretical predictions of matrix elements, or by considering
improved constraints on the EDMs in Table~\ref{tab:EDMexps} and bounds on the EDMs of additional systems, such as the proton or deuteron. 
Improvements on these three fronts are expected on the same timescale as the LHC Run III and the HL-LHC, for which the limits in Ref.~\cite{Bernlochner:2018opw} were derived.

\begin{figure}[t!]
	\center
	\includegraphics[width=0.7\linewidth]{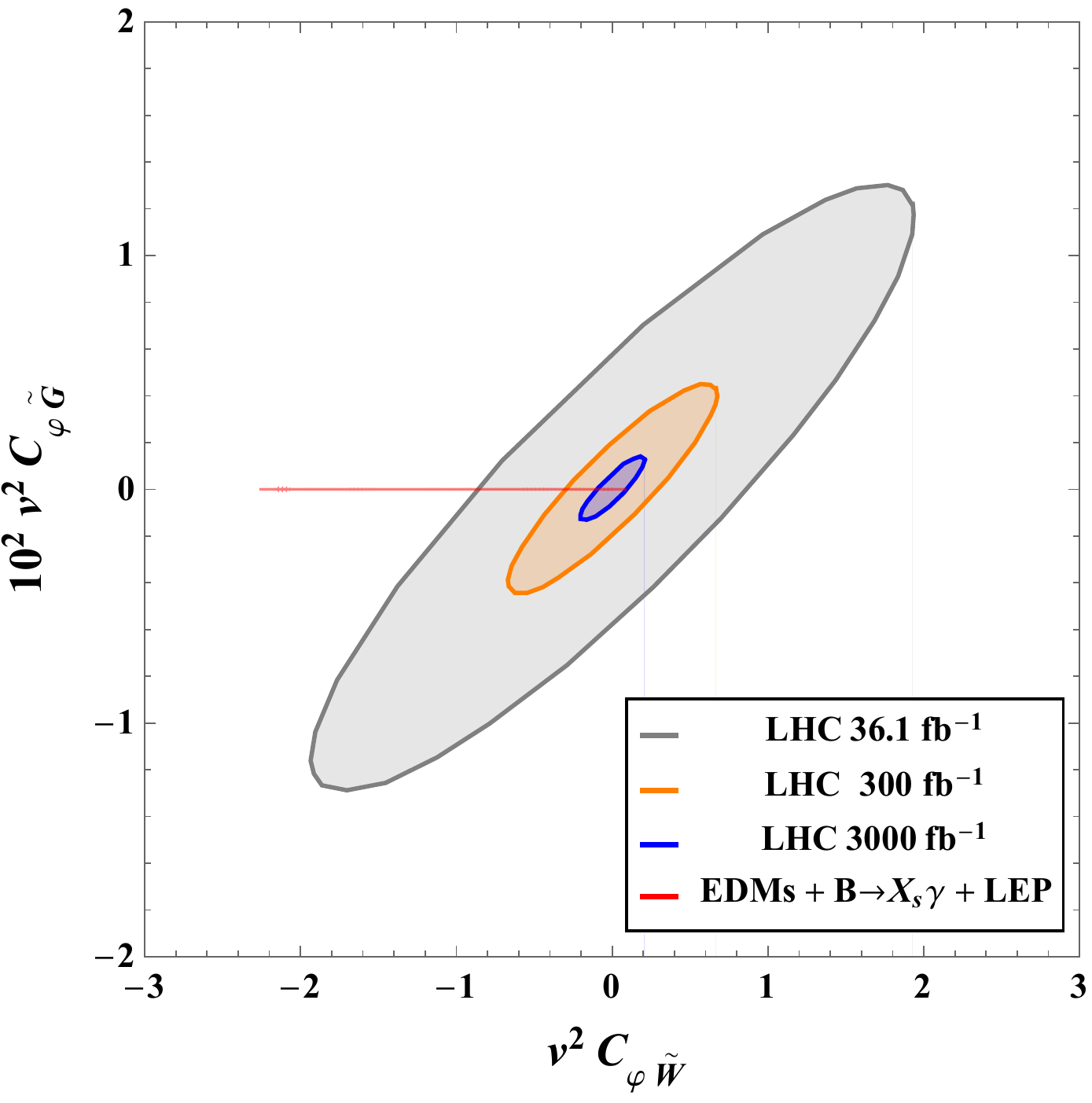}
	\includegraphics[width=0.7\linewidth]{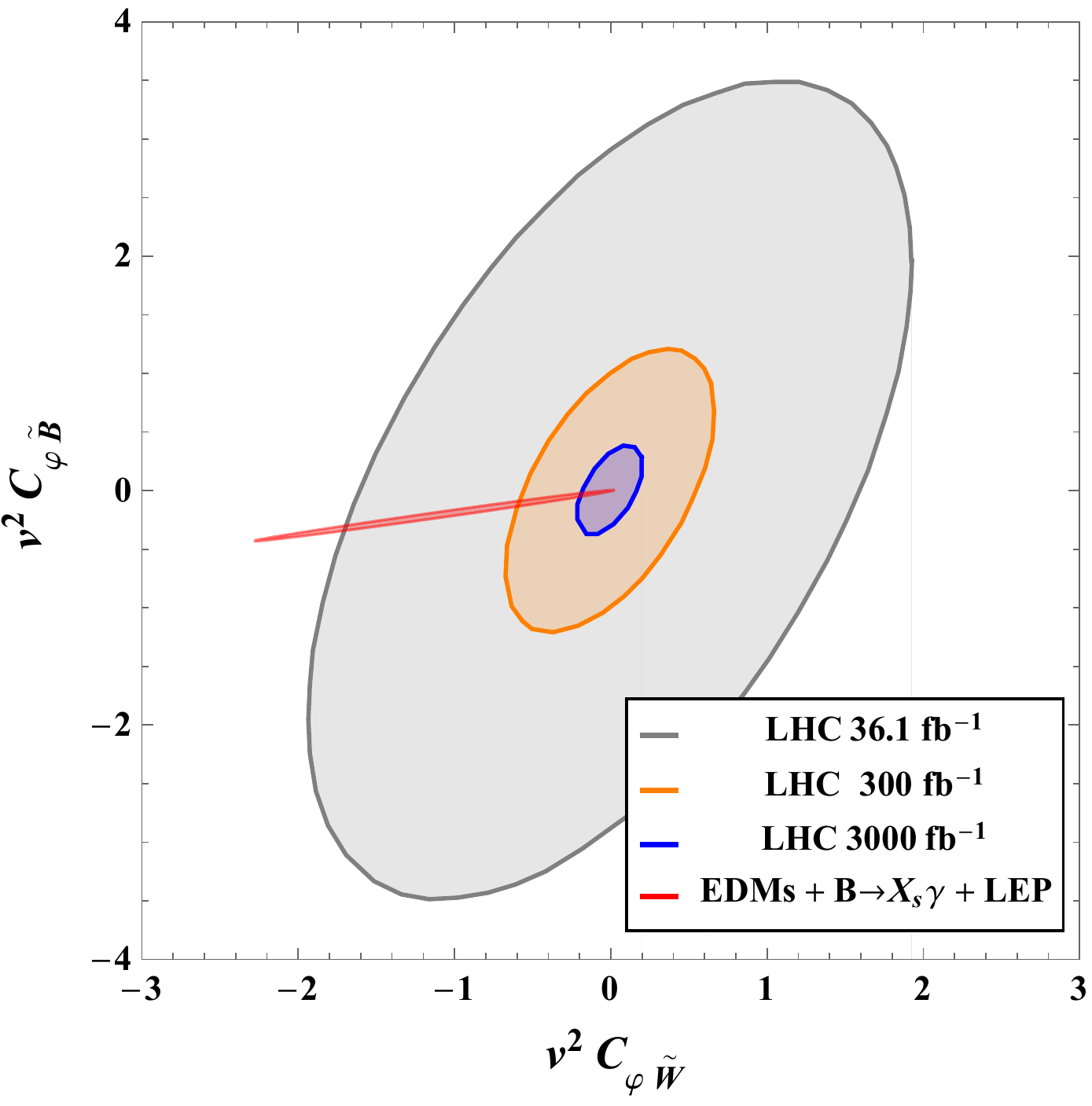}
	\caption{Projected $95\%$ C.L.\ constraints from EDM and $B\to X_s\gamma$  as well as collider signatures~\cite{Bernlochner:2018opw} in the $C_{\vp \tilde W}$--$C_{\vp \tilde G}$ and $C_{\vp \tilde W}$--$C_{\vp \tilde B}$ planes. The remaining couplings are marginalized over and the Gaussian strategy for the matrix elements is used. }
	\label{fig:collider}
\end{figure}

We  therefore consider improved determinations of the matrix elements that were set as targets for the future in Ref.~\cite{Chien:2015xha}. We assign 25\% uncertainties to the nucleon EDM induced by 
the $u$- and $d$-quark CEDMs, and  $50\%$ uncertainties on the nucleon EDM from $C_{\tilde G}$, 
the \CP-odd pion--nucleon couplings, and the nuclear structure matrix elements.
These uncertainty goals are
by no means unrealistic considering recent lattice and nuclear-theory efforts~\cite{Bhattacharya:2018qat,Syritsyn:2019vvt,Kim:2018rce,Rizik:2018lrz}, and in some cases have already
been attained~\cite{Dobaczewski:2018nim}.
On the experimental side, we assume $|d_n | < 1.0 \cdot 10^{-27}$ $e$ cm, which will be probed at the PSI and LANL neutron EDM experiments~\cite{Schmidt-Wellenburg:2016nfv,Ito:2017ywc},
and $|d_{\rm Ra}| < 10^{-27}$ $e$ cm, well within reach of the ANL radium EDM experiment~\cite{Bishof:2016uqx}.
On a longer timescale, storage ring searches of the EDMs of light ions have the potential to compete with the neutron EDM~\cite{Eversmann:2015jnk},
and we assume  $d_p, \, d_d < 1.0 \cdot 10^{-27}$ $e$ cm. For the \CPspace asymmetry in $B \rightarrow X_s \gamma$, Belle II
will be sensitive to sub-percent values, $|A_{B \rightarrow X_s \gamma}| < 4 \cdot 10^{-3}$~\cite{Kou:2018nap}.

A comparison of the projected limits of Ref.~\cite{Bernlochner:2018opw} to the combination of future EDM and $B\to X_s\gamma$ limits in the $C_{\vp \tilde W}$--$C_{\vp \tilde G}$
and $C_{\vp \tilde W}$--$C_{\vp \tilde B}$ planes is shown in Fig.~\ref{fig:collider} and in Table~\ref{Tab:Futbounds}. 
The non-zero central values for the low-energy curves are driven by the  LEP bound~\eqref{LEP} on $C_{\varphi \tilde W B}$, which deviates from zero by $\sim 2 \sigma$.
The gray, orange, and purple bands assume the proposed differential measurements in $pp\to h+2j$ have been performed on $36$, $300$, and $3000$  fb$^{-1}$ of integrated luminosity, respectively, 
while the red band shows the limits from low-energy experiments.
The figure shows that the collider observables could in principle probe the $C_{\vp \tilde W}$ and $C_{\vp \tilde B}$ couplings at a comparable level as the low-energy limits with $36$ and 
$3000$ fb$^{-1}$ of data, respectively, but become relevant only when delicate cancellations between different couplings occur. 
The low-energy constraints on the gluonic operator $C_{\vp \tilde G}$, are expected to be more stringent than the projected limits from the HL-LHC by roughly two orders of magnitude, see Table~\ref{Tab:Futbounds}. 

The  strong constraints that EDM experiments put on the parameter space 
will manifest themselves in correlations between observables at the LHC. For example, 
the electron EDM bound establishes correlations between $C_{\varphi \tilde W B}$, $C_{\varphi \tilde W}$,
and $C_{\varphi \tilde B}$, as can be seen from the lower panel in Fig.~\ref{fig:collider}.  An observation of large \CPspace violation in the Higgs--gauge sector, 
of the size of the right column in Table~\ref{Tab:Futbounds}, would then require a non-zero value for $C_{\vp \tilde WB}$. In such a scenario one would therefore expect 
large effects in diboson  production, induced by $C_{\vp \tilde WB}$, to be consistent with EDM experiments.

We can finally relax the assumption $C_{\tilde W, \tilde G}(\Lambda)=0$, and consider all the \CP-violating operators 
expected in the framework of universal theories. As argued above, the dominant EDM constraints  
are only sensitive to two linear combinations of the weak couplings $C_{\varphi \tilde B}$,
$C_{\varphi \tilde W}$, $C_{\varphi \tilde W B}$, and $C_{\tilde W}$, so that EDM experiments could, in total, 
provide four independent constraints on the six operators in Eq.~\eqref{HiggsGauge}. 
One possible strategy to close the open directions in parameter space relies on the \CPspace asymmetry in 
$B \rightarrow X_s \gamma$ and/or LEP observables, but of course complementary LHC measurements 
would  also provide the remaining two constraints. In either case, 
one again expects strong correlations between \CP-violating observables in the Higgs and weak boson sectors,
which illustrates the enormous potential of the low-energy probes in constraining the \CP-odd sector of universal theories.  

{\it Conclusions.}---In this Letter, we have analyzed the complementarity of LHC searches and low-energy experiments in
constraining or discovering \CPspace violation in Higgs--gauge interactions, in the context of universal theories. 
In particular, we studied {\it quantitatively} the impact of EDMs on the allowed parameter space.  
Our work shows that despite the loop suppression EDMs cannot be neglected (as in recent LHC analyses)---in fact in a single-operator analysis 
there is very little room for observing \CPspace violation in the Higgs sector at the LHC. 
In a global analysis, flat or weakly bound directions from low-energy constraints are still possible, defining which additional operator combinations are most useful 
to be constrained by the (HL-)LHC, via the observables considered in Refs.~\cite{Sirunyan:2019twz,Sirunyan:2019nbs,Aaboud:2017fye,Bernlochner:2018opw} 
and, potentially, EWPTs. 
Several lessons from our analysis extend beyond universal theories, 
where more \CP-violating effective couplings appear. In this case 
EDMs  enforce strong correlations among Higgs--gauge and other \CP-violating  couplings, 
which require either intricate cancellations and therefore insight on the new sources of \CPspace violation, 
or  strong bounds  on all the individual couplings.

\section*{Acknowledgments}
\begin{acknowledgments}
We thank Andrei Gritsan and Heshy Roskes for communication regarding Refs.~\cite{Sirunyan:2019twz,Sirunyan:2019nbs} and Uli Haisch for discussions.
This research is  supported by the U.S.\ Department of Energy,
 Office of Science, Office of Nuclear Physics, under contracts
DE-AC52-06NA25396, DE-FG02-00ER41132, and DE-SC0009919. AC is supported by a Professorship Grant (PP00P2\_176884) of the Swiss National Science Foundation. JdV is supported by the 
RHIC Physics Fellow Program of the RIKEN BNL Research Center.
\end{acknowledgments}

\bibliographystyle{h-physrev3} 
\bibliography{bibliography}

\section*{Supplemental Material}

{\it Renormalization group evolution.}---Here we provide additional details about the RG evolution and threshold corrections associated with the \CP-violating Higgs--gauge interactions. We also give detailed expressions for the low-energy observables used to set constraints on the Wilson coefficients of the operators in Eq.~\eqref{HiggsGauge} of the main text. These Higgs--gauge operators
induce the following operators through the diagrams shown in Fig.~\ref{diagrams} 
\begin{align}
\vL_{\rm EDMs} &=   \sum_{f=e,\mu,\tau,u,d,s,c,b,t}    \left(  C_\g^{(f)}  O_\g^{(f)}   +\text{h.c.} \right) \notag\\
&+   \sum_{f=u,d,s,c,b,t}  \left(  C_g^{(f)}  O_g^{(f)}   +\text{h.c.} \right)    + C_{\tilde G}  O_{\tilde G}\,, \notag\\ 
\vL_{b \to s}  &=   \frac{4G_F}{\sqrt{2}} V_{tb}V_{ts}^*\, C_7  O_7 \,,  
\label{eq:ExtendedEDM1}
\end{align}
with
\begin{align}
O^{(f)}_g &=  - \frac{g_s}{2}  m_f  \, \bar{q}^{(f)}_L  \sigma_{\mu \nu} G^{\mu \nu}_at^a q^{(f)}_R\,,\notag\\
O^{(f)}_{\gamma} &= - \frac{e Q_f}{2}  m_f  \, \bar{q}^{(f)}_L  \sigma_{\mu \nu} \left( F^{\mu \nu} -  t_w   Z^{\mu \nu} \right) q^{(f)}_R\,,\notag\\
O_7 &= \frac{e}{(4\pi)^2}m_b \bar q_L \sigma^{\mu\nu}F_{\mu\nu} b_R\,,
\end{align}
where $F^{\mu\nu}$ and $Z^{\mu\nu}$ are the photon and $Z$ field strengths, $t^a$ are the generators of $SU(3)_c$,  and $Q_f$ and $T^3_f$ represent the electric charge and third component of weak isospin. 
The flavor-changing gluonic dipole $O_8$ is further suppressed by $\alpha_{\rm em}/(4\pi)$ with respect to $C_7$, and can be safely neglected.
We use $e = -g s_w=-g'c_w$ and $t_w=s_w/c_w$, with $s_w = \sin \theta_w$, $c_w = \cos \theta_w$, and the weak mixing angle $\theta_w$. 
The Wilson coefficients of the dipole operators in Eq.~\eqref{eq:ExtendedEDM1}  are complex, we write $C_\alpha^{(f)} = c_\alpha^{(f)} + i \tilde c_\alpha^{(f)}$ with $c_\alpha^{(f)}$ and $\tilde c_\alpha^{(f)}$ real.

In addition to 
$B \rightarrow X_s \gamma$ as described by $\mathcal L_{b \to s}$,
$C_{\varphi \tilde W B}$ and $C_{\tilde W}$ induce $b \rightarrow d \gamma$ and $s \rightarrow d \gamma$ dipoles
via the same diagrams that contribute to $\mathcal L_{b \to s}$. The constraints from the direct \CPspace asymmetry in $B \rightarrow X_d \gamma$
and $B \rightarrow X_{s+d} \gamma$ are, respectively, weaker than and degenerate with  those from $B \rightarrow X_s \gamma$.
Similarly, constraints from \CP-violating observables in kaon physics, such as $K_L \rightarrow \pi^0 e^+ e^-$ 
are not competitive, and for these reasons we concentrate on $B \rightarrow X_s \gamma$.\footnote{Further complementary constraints could arise from $b\to s\ell\ell$ and $B_s\to\mu\mu$~\cite{Bobeth:2015zqa}.}
Altogether, we find the following matching conditions at $\mu_t=m_t\sim m_H$
\begin{align}
	\tilde c_\g^{(f)}&=\frac{\al_{\rm em}}{4\pi}\bigg[\bigg(\frac{3}{2}
	-\log \frac{m_h^2}{\mu_t^2}\bigg)\bigg(2\frac{T^3_f-2Q_f}{c_w\sq Q_f}C_{\vp \tilde B}\notag\\
	&-\frac{2T^3_f}{s_w\sq Q_f}C_{\vp\tilde W}+\frac{3T^3_f+t_w^2(2Q_f-T^3_f)}{s_w^2Q_f}C_{\vp \tilde WB}\bigg)\notag\\
	&-2\frac{T^3_f-2s_w\sq Q_f}{s_w\sq Q_f}\frac{m_Z\sq}{m_h^2-m_Z\sq}\log \frac{m_h\sq}{m_Z\sq}\notag\\
	&\times\left(t_w\sq C_{\vp \tilde B}-C_{\vp \tilde W}+\frac{1-t_w\sq}{2}C_{\vp \tilde WB}\right)\notag\\
	&-\frac{T^3_f}{s_w\sq Q_f} \left(2\log\frac{m_W\sq}{m_h\sq}  C_{\vp \tilde WB}-C_{\tilde W}\right)\bigg]\,,\notag\\
	\tilde c_g^{(q)}&=\frac{\al_s}{\pi}\left(\frac{3}{2}
	-\log \frac{m_h^2}{\mu_t\sq}\right) C_{\vp \tilde G}\,,\notag\\
	C_{\tilde G}(\mu_t^-) &= C_{\tilde G}(\mu_t^+) - \frac{\alpha_s(\mu_t^+)}{8 \pi} \tilde c^{(t)}_g (\mu_t^+)\,,\nn\\
C_7(\mu_t) &= i m_W^2\left(f(x_t)C_{\vp \tilde WB}+g(x_t)C_{\tilde W}\right)\,,
\label{eq:matching}
  \end{align}
with $x_t=m_t^2/m_W^2$ and loop functions~\cite{He:1993hx}
\begin{align}
 f(x) &= \frac{x}{(x-1)^3}(2-2x-x(x-3)\log x)\,,\notag\\
 g(x) &= \frac{x^3-x-2x^2\log x}{2(x-1)^3}\,.
\end{align}
The anomalous dimensions that determine the RG evolution can be extracted from the $\log \mu_t$ terms in Eq.~\eqref{eq:matching}. We have checked that these agree with results in Refs.~\cite{Dekens:2013zca,Alonso:2013hga}. 
The non-logarithmic terms in the matching relations depend on the regularization scheme.
Our result is valid in both naive dimensional regularization with anticommuting $\gamma_5$~\cite{Chanowitz:1979zu} and in the 't Hooft--Veltman scheme~\cite{tHooft:1972tcz}. 
In both cases the Levi-Civita tensor was considered as an external 4-dimensional object.
We use the RG evolution and the matching contributions to calculate the Wilson coefficients of the low-energy operators in  Eq.~\eqref{eq:ExtendedEDM1} at the EW scale. We take into account that $\tilde c_\g^{(q)}$, $\tilde c_g^{(q)}$, and $C_{\tilde G}$ renormalize under QCD~\cite{Weinberg:1989dx,Wilczek:1976ry,Braaten:1990gq,Degrassi:2005zd}. 
We then evolve the low-energy operators to the scale where QCD becomes non-perturbative, $\mu=\MQCD= 2$ GeV. At  the bottom, charm, and strange thresholds, the Weinberg operator obtains contributions analogous to the one in Eq.~\eqref{eq:matching}. The resulting fermion EDMs, CEDMs, the Weinberg operator, and $C_7$ at the scale $\mu=\MQCD$ are given in Table~\ref{Tab1}, where we assumed the initial scale $\mu_0= \Lambda=1$ TeV. 

As explained in the main text, the weak operators $C_{\varphi \tilde B}$, $C_{\varphi \tilde W}$, $C_{\varphi \tilde W B}$, and $C_{\tilde W}$ 
contribute to the fermion EDM $\tilde c_\gamma^{(f)}$ almost exclusively via two combinations,
proportional to $T^3_f$ and $Q_f$, respectively. Some  sensitivity to the linear combinations of weak operators that do not contribute to $\tilde c_{\gamma}^{(f)}$ 
arises when one considers additional EW loops or matching into \CP-violating semileptonic operators. 
The constraints are, however, weaker than the $B \rightarrow X_s \gamma$, LEP, and future collider constraints, so that we 
do not consider such effects here.

\begin{table}\small
\centering
\begin{tabular}{ccccccc}
\toprule
& $C_{\varphi \tilde B}$ & $C_{\varphi \tilde W}$ & $C_{\varphi \tilde W B}$ & $C_{\varphi \tilde G}$&$C_{\tilde G}$ &$C_{\tilde W}$\\
\colrule
$\tilde c^{(e)}_\gamma$ & $-1.4$ & $-1.5$ &  $3.3$& --& --&$0.14$\\
$\tilde c^{(u)}_\gamma$ & $-0.62$ & $-1.1$ &  $2.3$& $6.5$ & $-6.2$&$0.11$\\
$\tilde c^{(d)}_\gamma$ & $-0.31$ & $-2.2$ &  $4.0$& $6.5$&$-6.2$ &$0.22$\\
$\tilde c^{(q)}_g$ & -- & -- & -- & $10$ &$-15$&--\\
$C_{\tilde G}$       & -- & -- & -- & $-0.22$&$23$&--\\
$C_7$ &-- &-- &$-9.9\, v^2$ & --&-- &$1.7\, v^2$ \\
\botrule
\end{tabular}
\caption{Coefficients of the fermion EDMs, quark CEDMs, and Weinberg operator at $\mu = 2$ GeV in units of $10^{-2}$, assuming  $\mu_0 = \Lambda = 1$ TeV. The label $(d)$ denotes both the operators involving down and strange quarks, while the superscript $q$ in $\tilde c^{(q)}_g$ denotes $q = \{u,d,s\}$.}
\label{Tab1}
\end{table}

{\it Low-energy observables.}---In this section we provide explicit expressions for the low-energy observables used to constrain the Higgs--gauge operators. A more thorough discussion of all contributions and their uncertainties can be found in Ref.~\cite{Dekens:2018bci}. We begin with the constraint on the electron EDM. The most stringent constraint is set by the ACME collaboration~\cite{Baron:2013eja,Andreev:2018ayy} using the polar molecule ThO. In principle, the electron spin-precession frequency receives contributions from both the electron EDM and \CP-odd electron--nucleon interactions. The latter gets negligible contributions from the Higgs--gauge interactions under consideration, and we interpret the ThO measurement  as a limit on the electron EDM using the relation~\cite{Skripnikov,Fleig:2014uaa}
\begin{equation}
\omega_{\text{ThO}} =120.6(4.9)\mathrm{mrad}/\mathrm{s}\left(\frac{d_e}{10^{-27}\,e\,\mathrm{cm}}\right)\,,
\end{equation}
and the experimental limit $\omega_{\text{ThO}} < 1.3\,\mathrm{mrad}/\mathrm{s}$ at $90\%$ C.L.~\cite{Baron:2013eja,Andreev:2018ayy}. 

The neutron, $d_n$, and proton, $d_p$, EDMs are induced by
quark (C)EDMs and the Weinberg operator. Contributions from first-generation EDMs are
known with few percent accuracy~\cite{Gupta:2018lvp}. The contribution of the strange EDM has a larger uncertainty~\cite{Gupta:2018lvp, Alexandrou:2017qyt}. QCD sum-rule calculations determine the contributions from the  up- and down-quark CEDMs with roughly $50\%$ uncertainty, while the strange CEDM is assumed to vanish in the PQ scenario~\cite{Pospelov:2000bw,Lebedev:2004va,Pospelov:2005pr,Hisano:2012sc}. Contributions from the Weinberg operator are difficult to determine and  current estimates from QCD sum-rules~\cite{Demir:2002gg} and naive dimensional analysis~\cite{Weinberg:1989dx} have an $\Or(100\%)$ uncertainty
\begin{align}
d_n &=
-0.204(11)\,d_u+0.784(28)\,d_d\notag\\
&- 0.0028(17)\,d_s-0.55(28)\,e\,\tilde d_u\notag\\
&-1.10(55)\,e\,\tilde d_d +50(40)\,{\rm MeV}\,e\,g_sC_{\tilde G}\,,\notag\\
d_p&=
0.784(28)\,d_u-0.204(11)\,d_d\notag\\
&-0.0028(17)\,d_s +1.30(65)\,e\, \tilde d_u\notag\\
&+0.6(3)\,e\,\tilde d_d-50(40)\,{\rm MeV}\,e\,g_sC_{\tilde G}\,,
\end{align}
where $d_q =  Q_q m_q \tilde c_\g^{(q)}$ and  $\tilde d_q =  m_q \tilde c_g^{(q)}$.

EDMs of diamagnetic atoms and light nuclei receive contributions not only from the nucleon EDMs, but also from the \CP-violating nucleon--nucleon potential. For the operators under consideration, this potential is dominated~\cite{deVries:2012ab} by one-pion-exchange contributions involving the \CP-odd pion--nucleon ($\pi N$) vertices
\begin{equation}
\mathcal L_{\pi N} = \bar g_0 \bar N \boldtau\cdot \boldpi N +\bar g_1 \bar N \pi_3 N\,,
\end{equation}
in terms of the Pauli matrices $\boldtau$, the nucleon doublet $N = (p\,\,n)^T$, and the pion triplet $\boldpi$. 
The sizes of $\bar
g_{0,1}$ have been calculated with QCD sum rules~\cite{Pospelov:2001ys} 
\begin{align}
\bar g_0 &= 5(10)(m_u\tilde C^{(u)}_g + m_d\tilde C^{(d)}_g)\, \mathrm{fm}^{-1}\,,
\notag\\
\bar g_1 &= 20^{+40}_{-10}(m_u\tilde C^{(u)}_g- m_d\tilde C^{(d)}_g)\,\, \mathrm{fm}^{-1}\,.
\end{align}
In combination with \CP-even nuclear forces and currents, the nucleon EDMs and the \CP-odd $\pi N$ interactions can be used to calculate the EDMs of light nuclei. 
In particular, the EDM of the deuteron is given by~\cite{Bsaisou:2014zwa}
\begin{equation}
d_{D} =
0.94(1)(d_n + d_p) + 0.18(2) \,\bar g_1 \,e \,{\rm fm} \, .
\end{equation}

The EDM of the diamagnetic atom ${}^{199}$Hg gets contributions from both nuclear and leptonic \CP-odd interactions. For our purposes we use~\cite{Dmitriev:2003sc,deJesus:2005nb,Ban:2010ea,Dzuba:2009kn,Engel:2013lsa,Sahoo:2016zvr,Fleig:2018bsf}
\begin{align}
 d_{\rm Hg}&= -2.1(5)
\Ex{-4}\bigg[1.9(1)d_n +0.20(6)d_p\notag\\
&+\bigg(0.13^{+0.50}_{-0.07}\,\bar g_0 +
0.25^{+0.89}_{-0.63}\,\bar g_1\bigg)e\, {\rm fm}\bigg]\notag\\
&+0.012(12) d_e \,,
\end{align}
neglecting semileptonic interactions~\cite{Yamanaka:2017mef,Engel:2013lsa,Yanase:2018qqq} that receive small contributions from the Higgs--gauge operators.
Due to octopole deformations, the EDM of ${}^{225}$Ra is dominated by the pion-exchange contributions. We use~\cite{Dobaczewski:2018nim}
\begin{equation}
d_{\mathrm{Ra}} = -7.7(8)\Ex{-4}[-2.5(7.6) \,\bar g_0 
+ 63(38)\,\bar g_1] e\, {\rm fm}\,.
\end{equation}
Although it has little affect in our analysis, for completeness we provide the following expression for $^{129}$Xe~\cite{Dzuba:2009kn,Engel:2013lsa}
\begin{align}
d_{\mathrm{Xe}} &= 0.33(5)\Ex{-4}\Big[-0.32(2) d_n+0.0061(10) d_p\notag 
\\
&+\left(-0.10^{+0.04}_{-0.53}\,\bar g_0 
-0.08^{+0.04}_{-0.55}\,\bar g_1\right) e\, {\rm fm}\,\Big]\,.
\end{align}

Finally, we employ the expressions of Ref.~\cite{Benzke:2010tq} for the \CPspace asymmetry in $B\to X_s\g$
\begin{align}
\frac{A_{CP}}{\pi}&\equiv \frac{1}{\pi}\frac{\Gamma(\bar B\to X_s\g)- \Gamma(B\to X_{\bar s}\g)}{\Gamma(\bar B\to X_s\g)+ \Gamma(B\to X_{\bar s}\g)}\notag\\
&\left[\left(\frac{40}{81}-\frac{40}{9}\frac{\Lambda_c}{m_b}\right)\frac{\al_s}{\pi}+\frac{\Lambda_{17}^c}{m_b}\right]{\rm Im}\frac{C_2}{C^{\rm tot}_7}\notag\\
&-\left(\frac{4\al_s}{9\pi}+4\pi\al_s\frac{\Lambda_{78}}{3m_b}\right){\rm Im}\frac{C_8}{C^{\rm tot}_7}\notag\\
&-\left(\frac{\Lambda_{17}^u-\Lambda_{17}^c}{m_b}+\frac{40}{9}\frac{\Lambda_c}{m_b}\frac{\al_s}{\pi}\right){\rm Im}\left(\epsilon_s\frac{C_2}{C^{\rm tot}_7}\right)\,,
\end{align}
where $\epsilon_s=\frac{V_{us}^*V_{ub}}{V_{ts}^*V_{tb}}$ and we use 
\begin{align}
\Lambda_{17}^u &= [-0.33,\, 0.525]\GeV\,,\notag\\ 
\Lambda_{17}^c &= [-0.009,\, 0.011]\GeV\,,\notag\\
\Lambda_{78} &= [-0.017,\, 0.19]\GeV\,.
\end{align}
The Wilson coefficients are given by 
\begin{align}
C_2&=C_2^{\rm SM}(\mu_b) = 1.204\,,\notag\\ 
C_8&=C_8^{\rm SM}(\mu_b) = -0.175\,,\notag\\
C^{\rm tot}_7&=C_7^{\rm SM}(\mu_b)+C_7(\mu_b) = -0.381+C_7(\mu_b)\,.
\end{align}

{\it CMS limits.}---Refs.~\cite{Sirunyan:2019twz,Sirunyan:2019nbs} 
consider limits on \CP-violating effective couplings that affect the production of a Higgs boson via vector-boson fusion (VBF), with the Higgs subsequently decaying into $\tau\tau$ or four leptons. The observables discussed in Ref.~\cite{Sirunyan:2019twz,Sirunyan:2019nbs}
are mostly sensitive to the modification of the $h Z Z$ vertex, which affects both VBF and $h\rightarrow 4l$, and of the $h W W$ 
vertex, which affects VBF. The bounds in~\cite{Sirunyan:2019twz,Sirunyan:2019nbs} are expressed in terms of the anomalous coupling $a_3$, parameterizing  \CP-violating contributions to 
the $h \rightarrow ZZ$ vertex, and $r_{a3}$, the ratio of the $CP$-violating $hWW$ and $hZZ$ vertices. 
The coefficients of the SMEFT operators defined in Eq.~(1) of the manuscript can be mapped onto the effective couplings $a_3$ and $r_{a3}$ 
as follows
\begin{align}\label{eq.a3}
\frac{a_3}{a_1} &=  \frac{ g^2}{c_w^2}\\
&\times\left( c_w^4 (v^2 C_{\varphi \tilde W}) 
+ s_w^4 (v^2 C_{\varphi \tilde B}) + c_w^2 s_w^2 (v^2 C_{\varphi \tilde W B}) 
\right)\,, \notag \\
r_{a3} &=  \frac{c_w^2 (v^2 C_{\varphi \tilde W}) }{c_w^4 (v^2 C_{\varphi \tilde W}) 
+ s_w^4 (v^2 C_{\varphi \tilde B}) + c_w^2 s_w^2 (v^2 C_{\varphi \tilde W B})}\,,\notag
\end{align}
where $a_1$ denotes the SM $h ZZ$ coupling.
In the analyses of Refs.~\cite{Sirunyan:2019twz,Sirunyan:2019nbs},  $r_{a3}$ is, for convenience, set to one. 
The observed 95\% C.L.\ constraints in Ref.~\cite{Sirunyan:2019nbs} are however insensitive to the value of $r_{a3}$ ~\cite{CMSprivate},
and can thus be interpreted as a bound on $a_3/a_1$ as given in Eq.~\eqref{eq.a3}.

Ref.~\cite{Sirunyan:2019nbs} presents a strong $68\%$ C.L.\ limit, which is dominated by corrections to VBF, and almost reaches the 2$\sigma$ level.
In this case, the results in Ref.~\cite{Sirunyan:2019nbs} depend on the choice of $r_{a3}$, but, being dominated by VBF, the constraints obtained with  $r_{a3}= 1$ can be 
converted into constraints with arbitrary $r_{a3}$~\cite{Sirunyan:2019twz,CMSprivate}. As an estimate of the bounds  
that can be reached in the near future, we quote the expected 95\% C.L.\ limit in Ref.~\cite{Sirunyan:2019nbs}, which gives 
\begin{equation}
| v^2 C_{\varphi \tilde{B}}| < 6\,, \quad
| v^2 C_{\varphi \tilde{W}}| < 0.22\,, \quad
| v^2 C_{\varphi \tilde W  B} | < 1.8\,.
\end{equation}

\end{document}